\documentclass[pdflatex,sn-nature]{sn-jnl}% Style for submissions to Nature Portfolio journals

\usepackage{graphicx}%
\usepackage{multirow}%
\usepackage{amsmath,amssymb,amsfonts}%
\usepackage{amsthm}%
\usepackage{mathrsfs}%
\usepackage[title]{appendix}%
\usepackage{xcolor}%
\usepackage{textcomp}%
\usepackage{manyfoot}%
\usepackage{booktabs}%
\usepackage{algorithm}%
\usepackage{algorithmicx}%
\usepackage{algpseudocode}%
\usepackage{listings}%
\usepackage{colortbl}

\usepackage{wrapfig}
\usepackage{comment}
\usepackage{subcaption}
\definecolor{tabhead}{RGB}{214, 213, 193}
\definecolor{tabc1}{RGB}{250, 250, 246}
\definecolor{tabc2}{RGB}{236, 235, 226}

\theoremstyle{thmstyleone}%
\theoremstyle{thmstyletwo}%

\theoremstyle{thmstylethree}%

\begin{document}

\title[Article Title]{Reservoir-enhanced Segment Anything Model for Subsurface Diagnosis}

%%=============================================================%%
%% GivenName	-> \fnm{Joergen W.}
%% Particle	-> \spfx{van der} -> surname prefix
%% FamilyName	-> \sur{Ploeg}
%% Suffix	-> \sfx{IV}
%% \author*[1,2]{\fnm{Joergen W.} \spfx{van der} \sur{Ploeg} 
%%  \sfx{IV}}\email{iauthor@gmail.com}
%%=============================================================%%

\author[1]{\fnm{Xiren} \sur{Zhou}}\email{zhou0612@ustc.edu.cn}

\author[1]{\fnm{Shikang} \sur{Liu}}\email{skliu00@mail.ustc.edu.cn}

\author[1]{\fnm{Xinyu} \sur{Yan}}\email{xinyuyan@mail.ustc.edu.cn}

\author[1]{\fnm{Yizhan} \sur{Fan}}\email{fyz666@mail.ustc.edu.cn}

\author[1]{\fnm{Xiangyu} \sur{Wang}}\email{sa312@ustc.edu.cn}

\author[2]{\fnm{Yu} \sur{Kang}} \email{kangduyu@ustc.edu.cn}

\author[3]{\fnm{Jian} \sur{Cheng}} \email{jiancheng@tsinghua.org.cn}

\author*[1]{\fnm{Huanhuan} \sur{Chen}}\email{hchen@ustc.edu.cn}
% \equalcont{These authors contributed equally to this work.}

\affil*[1]{\orgdiv{School of Computer Science and Technology}, \orgname{University of Science and Technology of China}, \orgaddress{\street{No. 96 Jinzhai Road}, \city{Hefei}, \postcode{230027}, \state{Anhui}, \country{China}}}

\affil*[2]{\orgdiv{Department of Automation}, \orgname{University of Science and Technology of China}, \orgaddress{\street{No. 443, Huangshan Road}, \city{Hefei}, \postcode{230026}, \state{Anhui}, \country{China}}} 

\affil*[3]{\orgdiv{Research Institute of Mine Artificial Intelligence}, \orgname{Chinese Institute of Coal Science}, \orgaddress{\street{No. 5, Qingniangou Road}, \postcode{100013}, \state{Beijing}, \country{China}}}

% \affil[2]{\orgdiv{Department}, \orgname{Organization}, \orgaddress{\street{Street}, \city{City}, \postcode{10587}, \state{State}, \country{Country}}}

% \affil[3]{\orgdiv{Department}, \orgname{Organization}, \orgaddress{\street{Street}, \city{City}, \postcode{610101}, \state{State}, \country{Country}}}

%%==================================%%
%% Sample for unstructured abstract %%
%%==================================%%

\abstract{

Urban roads and infrastructure, vital to city operations, face growing threats from subsurface anomalies like cracks and cavities. Ground Penetrating Radar (GPR) effectively visualizes underground conditions employing electromagnetic (EM) waves; however, accurate anomaly detection via GPR remains challenging due to limited labeled data, varying subsurface conditions, and indistinct target boundaries. Although visually image-like, GPR data fundamentally represent EM waves, with variations within and between waves critical for identifying anomalies. Addressing these, we propose the Reservoir-enhanced Segment Anything Model (Res-SAM), an innovative framework exploiting both visual discernibility and wave-changing properties of GPR data. Res-SAM initially identifies apparent candidate anomaly regions given minimal prompts, and further refines them by analyzing anomaly-induced changing information within and between EM waves in local GPR data, enabling precise and complete anomaly region extraction and category determination. Real-world experiments demonstrate that Res-SAM achieves high detection accuracy ($>$ 85\%) and outperforms state-of-the-art. Notably, Res-SAM requires only minimal accessible non-target data, avoids intensive training, and incorporates simple human interaction to enhance reliability. Our research provides a scalable, resource-efficient solution for rapid subsurface anomaly detection across diverse environments, improving urban safety monitoring while reducing manual effort and computational cost.

}


%%\pacs[JEL Classification]{D8, H51}

%%\pacs[MSC Classification]{35A01, 65L10, 65L12, 65L20, 65L70}

\maketitle

\section{Introduction}\label{sec1}

As urbanization accelerates globally, rapid expansion of critical infrastructure increasingly encounters threats from subsurface hazards, such as cavities, soil loosening, and structural voids \cite{liu2025urban}. These underground anomalies pose significant risks, potentially leading to severe subsidence that endangers roads, buildings, and public safety \cite{habitat2022world}. In China alone, incomplete statistics indicate that over 1,000 underground disasters occur annually, ranging from minor ground settlements to catastrophic collapses, which result in substantial economic losses and risks to human life \cite{xie2023narrative}. Detecting and locating these hidden anomalies reliably and efficiently thus becomes critical.

\begin{figure}[htbp]
    \centering
    \begin{subfigure}[b]{0.38\textwidth}
        \centering
        \includegraphics[height=1.03in]{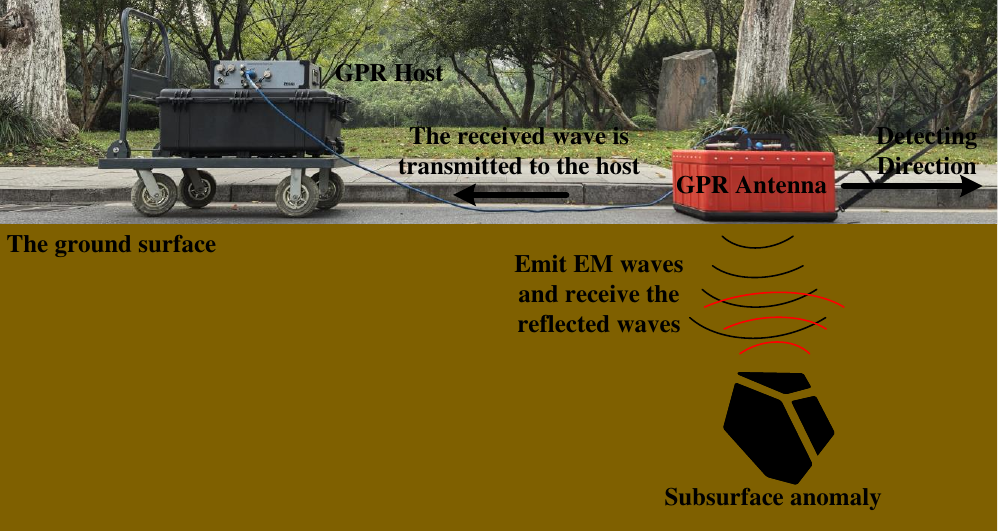}
        \caption{GPR Equipment}
        \label{GPR_Detail}
    \end{subfigure}
    \hfill
    \begin{subfigure}[b]{0.58\textwidth}
        \centering
        \includegraphics[height=1.05in]{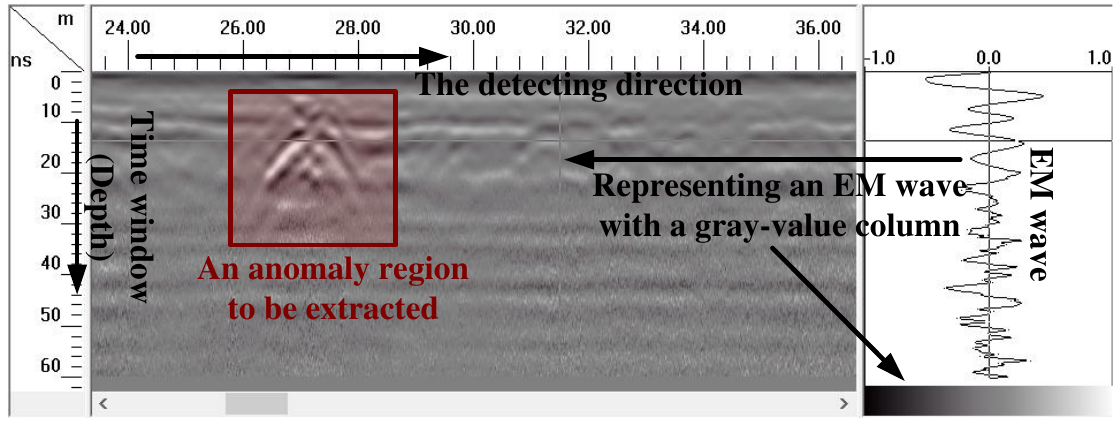}
        \caption{GPR B-scan Data}
        \label{abscan}
    \end{subfigure}
    \caption{\textbf{(a)} GPR typically consists of a host and an antenna. The antenna emits EM waves into the ground, receiving reflections from subsurface anomalies. These signals are transmitted to the host for processing and visualized as B-scan images. \textbf{(b)} Left side shows a B-scan with horizontal and vertical axes representing detection position and depth-related time windows. Right side shows a waveform that correlates gray values with wave amplitude variations. In this paper, ``GPR data'' specifically refers to GPR B-scan data.}
    \label{fig:SAM_example}
\end{figure}

Ground Penetrating Radar (GPR) has emerged as a crucial non-invasive geophysical tool for detecting underground facilities and anomalies \cite{zhou2022mapping,shen2023quality}. As illustrated in Fig. \ref{GPR_Detail}, GPR emits high-frequency electromagnetic (EM) waves into the ground and captures their reflections \cite{conyers2002ground}. The received signals could be collected horizontally as gray values in a B-scan format (Fig. \ref{abscan}), effectively visualizing underground situations \cite{zhou2018automatic}. Nevertheless, with expanding detection, underground diagnosis requires precise identification and localization of anomaly regions in GPR data, a laborious and time-consuming process when performed manually \cite{liu2024D3}. 

% Traditional image or signal feature extraction methods \cite{lee2014automatic,ye2014board,todkar2017detection,noreen2017using} assist in GPR data segmenting and categorizing, yet struggle with the variability in subsurface media and anomalies due to differences in composition, size, and surrounding environment \cite{ozkaya2020gpr}.  Shift from image processing, Deep learning (DL) techniques, primarily Convolutional Neural Networks (CNNs), have also been utilized to identify objects in GPR data \cite{9103247,hou2021deep,liu2021gprinvnet,liu2023automatic,wei2024tgrs}. However, practical challenges remain significant: the scarcity of multi-type labeled data, especially in a given undetected area, typically yields intialiazing datasets dominated by ``non-target'' samples free from any underground diseases like cavities/looseness; the diverse underground conditions hamper the generalization ability of learning methods, restricting their adaptability to diverse areas; the complex and extensive parameterization of DL networks necessitates heightened computational resources and data, failing to be efficiently optimized given limited data \cite{zhou2023c3modelspace}.

Traditional image or signal feature extraction methods \cite{lee2014automatic,ye2014board,todkar2017detection,noreen2017using} assist in GPR data segmenting and categorizing, yet struggle with the variability in subsurface media and anomalies due to differences in composition, size, and surrounding environment \cite{ozkaya2020gpr}.  Shift from image processing, Deep learning (DL) techniques, primarily Convolutional Neural Networks (CNNs), have also been utilized to identify objects in GPR data \cite{9103247,hou2021deep,liu2021gprinvnet,liu2023automatic,wei2024tgrs}. Nevertheless, DL approaches still face practical challenges. Firstly, there is a notable scarcity of multi-type labeled anomaly data, resulting in initial datasets heavily skewed toward abundant non-target samples (i.e., normal subsurface conditions without anomalies such as cavities or looseness). Secondly, the variability in underground conditions restricts the generalization capabilities of DL models, limiting their adaptability across diverse detection environments. Thirdly, the complexity and extensive parameterization of deep neural networks demand substantial computational resources, complicating efficient optimization when data availability is limited \cite{zhou2023c3modelspace}. Apart from the above, end-to-end DL methods indiscriminately process large volumes of normal data prevalent in practical GPR surveys, thereby consuming excessive computational resources and processing time. The absence of efficient human intervention may also reduce the reliability and interpretability of anomaly detection outcomes.

Recent advances in foundational vision models, such as the Segment Anything Model (SAM) \cite{kirillov2023segment}, have shown potential in anomaly segmentation, eliminating the need for additional data collection or training for specific tasks. SAM segments visual objects supported with minimal and user-friendly manual prompts, suggesting promise for detecting visually discernible underground anomalies. It reduces manual effort while improving result reliability through human intervention \cite{ma2024segment,oh2024llm,hekrdla2025optimized,debiasi2025phage}. Nonetheless, directly applying SAM in GPR data is hindered by the modality gap between optical images and EM waves. Despite being image-like as Fig. \ref{abscan}, GPR data essentially represent EM waves. Although visually distinguishable, local variations caused by subsurface anomalies are expressed as dual-directional dynamic changes, both within and between these waves, manifesting as gradual waveform alterations rather than distinct object boundaries \cite{chen2022underground}. Consequently, SAM is prone to over-segmentation or incomplete anomaly delineation, underscoring a key challenge: the dual-directional changing information in GPR data remains underutilized in existing frameworks \cite{zhou2024learning}.

To address these challenges, we propose the Reservoir-enhanced Segment Anything Model (Res-SAM), a framework that combines reservoir computing with interactive segmentation techniques to accurately identify and categorize anomalies in GPR data. Res-SAM uniquely integrates visual discernibility of anomalies with the subtle changes within and between EM waves inherent in local GPR data. Initially guided by simple manual click prompts, candidate anomaly regions are efficiently localized, which are further refined by analyzing waveform changing information in local GPR data via a specifically designed reservoir computing approach, enabling precise and complete anomaly region extraction and category determination. Significantly reducing dependence on extensive labeled data, Res-SAM operates effectively using minimal, easily accessible non-target samples and avoids resource-intensive iterative training. Moreover, it incorporates intuitive human interaction, enhancing practical reliability and interpretability. Experiments demonstrate that Res-SAM substantially outperforms existing approaches, highlighting its potential as a scalable, resource-efficient solution for rapid, accurate subsurface anomaly detection across diverse urban environments.

\section{Results}\label{results}
\subsection{Framework Overview}

Detecting underground anomalies using GPR faces practical challenges. First, anomaly involves unknowns, thus, there is a notable shortage of labeled anomaly data. Acquiring anomaly samples is costly, resulting in highly imbalanced datasets dominated by normal (non-target) samples. This imbalance severely restricts the training process of data-driven approaches. Second, the high variability of subsurface environments, including diverse soils and structures, leads to significant differences in EM wave characteristics. Such variability impedes the generalization capability, causing approaches trained on one scenario to likely perform poorly in others. Third, underground anomalies typically present as gradual changes in EM wave reflections rather than sharp, distinct boundaries. This gradual transition complicates precise anomaly localization, making conventional methods designed for visual images inadequate for wholly and accurately delineating anomaly boundaries in real-world applications.

\begin{figure}[htpb]
	\centering
	\includegraphics[width=\linewidth]{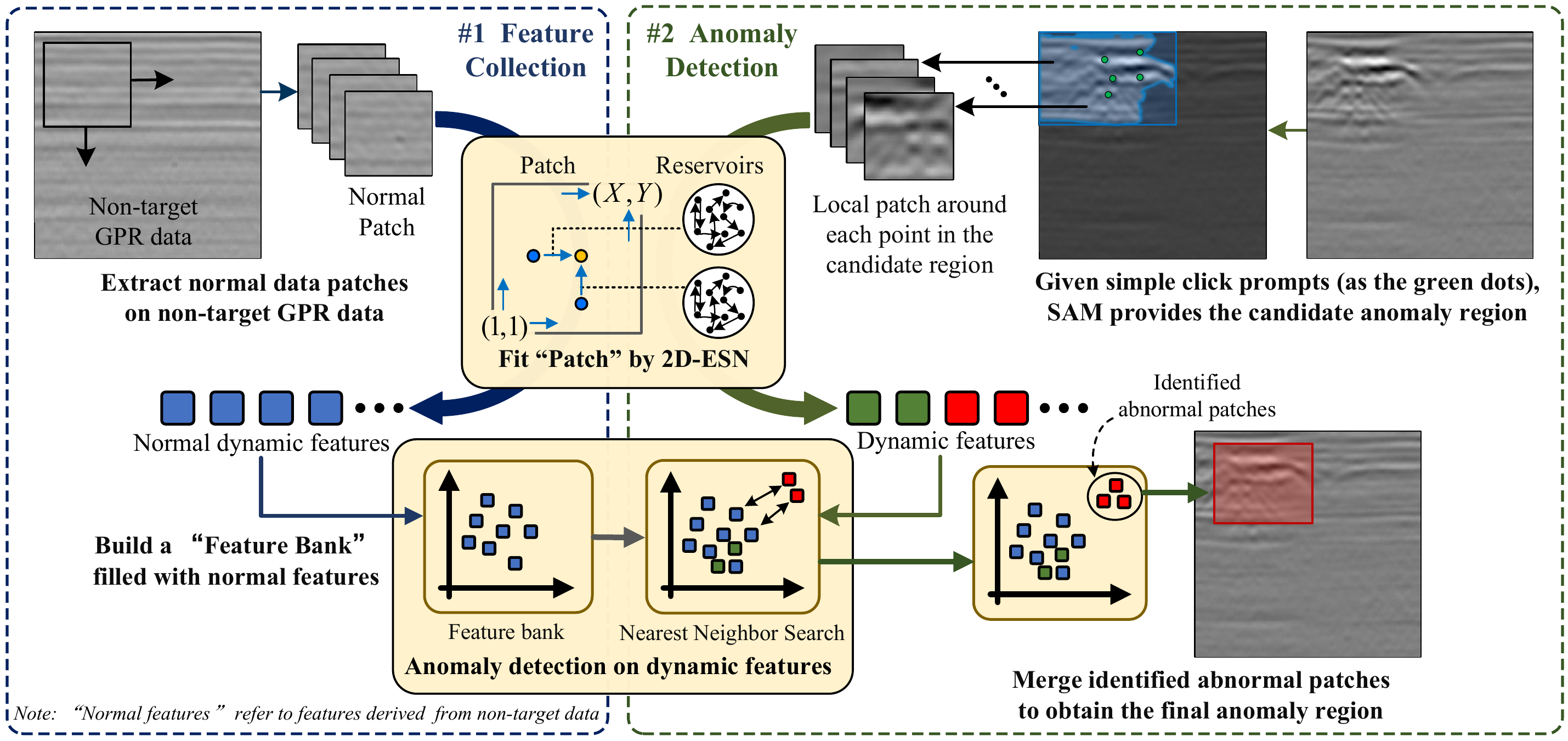}
	\caption{Res-SAM framework includes two phases: Feature Collection and Anomaly Detection. In the Feature Collection phase (\#1), normal data patches are extracted from non-target GPR data and fitted using 2D-ESN, with the derived dynamic features collected into a feature bank. In the Anomaly Detection phase (\#2), SAM initially identifies candidate anomaly regions. Local patches around each point in the region are fitted similarly using 2D-ESN, deriving features that are compared with the feature bank, with anomaly-associated patches identified and further merged into the final anomaly region.
    }
	\label{fig:framework}
\end{figure}

The proposed Res-SAM leverages minimal non-target training data, readily obtainable from the detection area, and avoids computationally intensive training, enabling rapid deployment in unexplored environments. It synergizes visual discernibility with local dynamic variations in GPR data caused by underground anomalies. Recognizing that practical subsurface anomaly detection necessitates clearly delineating entire affected regions, Res-SAM provides rectangular outlines of anomalies along with their categorization. As depicted in Fig. \ref{fig:framework}, Res-SAM operates in two main phases: ``Feature Collection'' and ``Anomaly Detection''.

In the Feature Collection phase, local patches are extracted from non-target GPR data and fitted individually using the Dual-Directional Echo State Network (2D-ESN). Acknowledging the vertical continuity of EM waves and horizontal correlations due to the consistency of the subsurface medium, effective changing information exists both vertically and horizontally within GPR data. 2D-ESN, a Reservoir Computing Network that incorporates two reservoirs in its hidden layer, connects each data point within the patch to its neighbors in both vertical and horizontal directions. Fitting each patch with 2D-ESN adequately captures the local data-inherent dual-directional changing information, resulting in a compact fitted readout model serving as ``dual-directional dynamic features'' or simply ``features'' of the original patch. These features, derived from patches on non-target data and denoted as ``normal features'', collectively represent local changing information in non-target data and are stored in a ``Feature Bank''.

In the Anomaly Detection phase, SAM is first introduced to identify initial candidate anomaly regions, guided by simple click prompts (typically single-digit number left- and right-mouse clicks). However, anomalies in GPR data often have indistinct boundaries due to EM wave reflection, diffraction, and attenuation, necessitating further refinement. To this end, we focus on local changing information within GPR data caused by underground anomalies, aiming to precisely adjust the candidate region. Specifically, for each point within the candidate region, we extract a local patch centered on this point and fit this patch using 2D-ESN. The obtained dynamic features are compared with the normal features in the feature bank. Patches exhibiting significantly different dynamics (high anomaly likelihood) indicate anomalies; overlapping anomaly patches are then merged to precisely delineate anomaly regions. Owing to the similar changing information in GPR data, similar subsurface structures result in comparable dynamic features, while those from different structures exhibit significant variances, reflecting their distinct changing information along and among EM waves. Therefore, the confirmed anomaly regions undergo another round of 2D-ESN fitting to derive category-specific dynamic features, enabling anomaly categorization.

\subsection{Dataset acquisition}

The utilized GPR data was collected along cement and asphalt surfaces, the most prevalent urban road types, using a GSSI GPR system (Fig. \ref{GPR_Detail}). Raw GPR data was initially acquired and preprocessed to reduce noise and enhance subsurface feature visibility. Surface reflections were removed, followed by electromagnetic noise reduction using a median filter \cite{olhoeft2000maximizing}, and time-varying gain compensation to address signal attenuation \cite{strange2002signal}. 
The GPR B-scan data was segmented into 369×369 pixel frames. Vertically, each frame spans a 64 ns time window, corresponding to a depth of about 6 meters, while horizontally covering approximately 15 meters. Frames containing subsurface anomalies like cavities or structures such as pipelines were referred to as ``abnormal'', while those without such objects were designated as ``non-target'' frames. 

 \begin{wrapfigure}{r}{0.42\linewidth} 
    \centering
    %\vspace{-1.5em}
    \includegraphics[width=\linewidth]{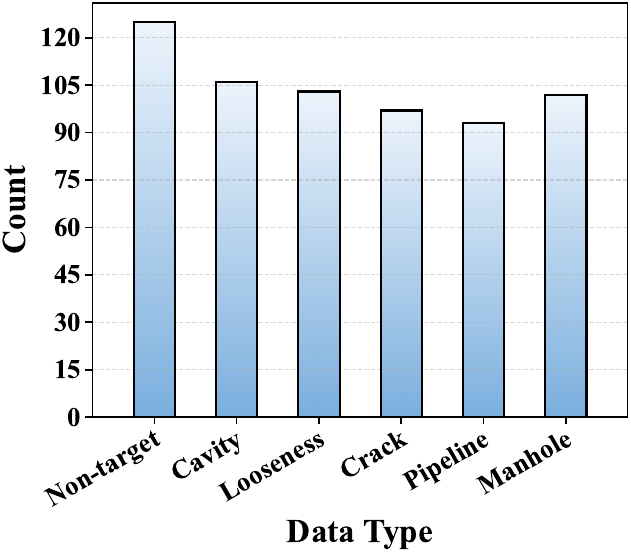}
    \caption{Distribution of the collected GPR data frames: type and amount.}
    \label{GPR_type_amount}
    \vspace{-1.0em}
\end{wrapfigure}

Similar to other underground anomaly detection tasks, the collected data is unbalanced, with non-target data far outnumbering anomaly-associated ones. We randomly subsampled non-target frames to achieve approximate class balance. This yielded a final dataset of 626 frames partitioned into six categories. The amount of frames across categories is summarized in Fig. \ref{GPR_type_amount}.

Our dataset encompasses three primary subsurface anomaly types: cavities, looseness, and cracks, differing in size and geophysical characteristics, also the most common ones in real-world detection. Cavities are air pockets under roads, contributing to urban road subsidence. Their low dielectric constant causes strong EM wave reflections, leading to multiple reflections and diffraction effects. Looseness, or ``loose'' areas, are regions with greater porosity and lower density compared to the surrounding soil. Their lesser cohesion can cause instability and voids. EM waves show chaotic reflections in GPR data for such soil. Cracks, characterized by horizontal separations within underground strata, are typically air-filled and cause sudden waveform disruptions in captured waves. In addition to these anomalies, our dataset includes GPR data of two common underground structures: pipelines and manhole covers (shortened as ``manhole''), which are also considered ``abnormal'' and need to be identified.

\begin{figure}[tbp]
    \centering
    \includegraphics[width=0.9\linewidth]{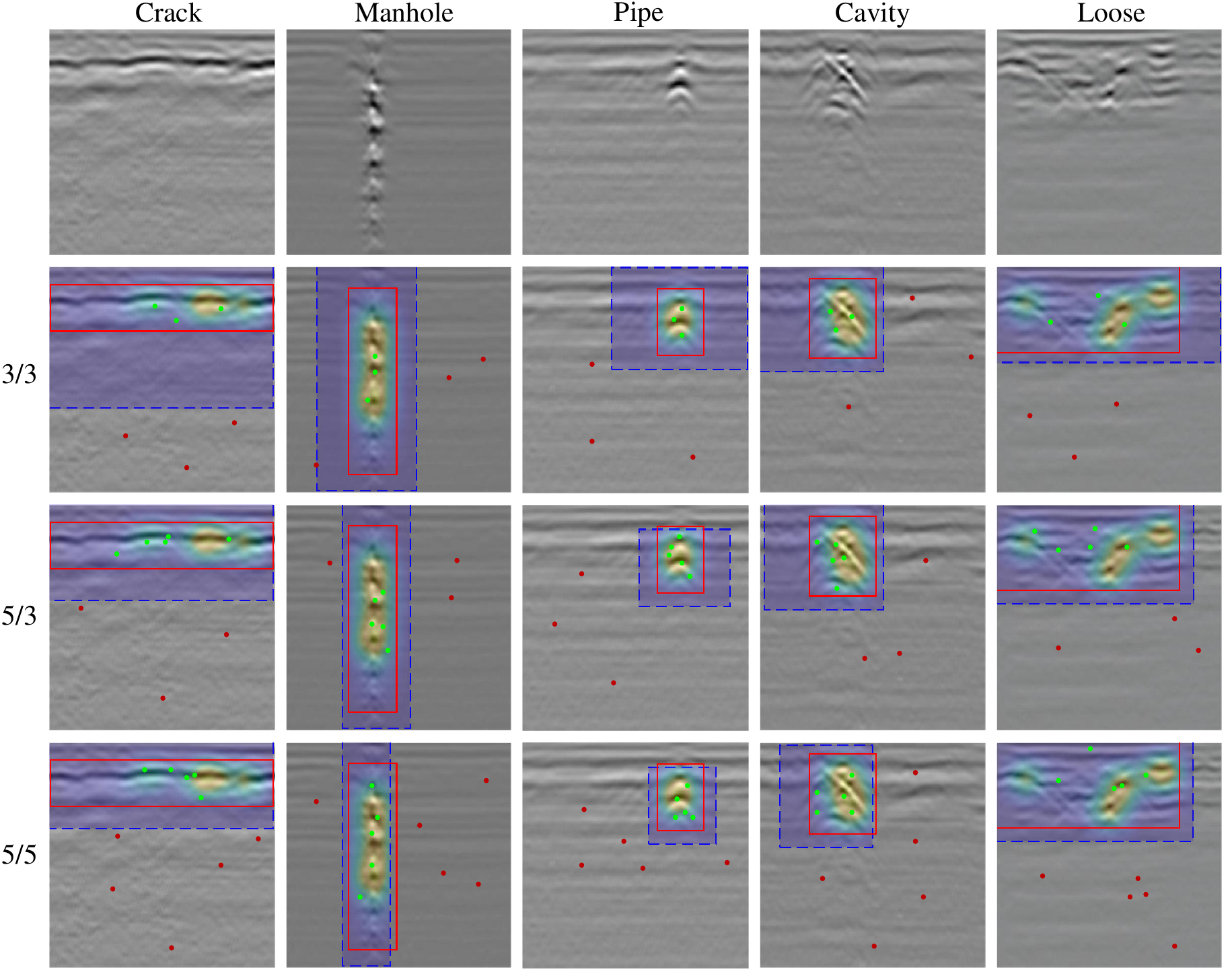}
    \caption{The candidate and final anomaly regions identified via Res-SAM. Each column represents a different type of anomaly (Cavity, Crack, Loose soil, Manhole, and Pipe). The first row shows the original GPR data. The blue dotted boxes highlight the candidate regions determined by SAM. The heatmaps overlaying the images show the anomaly likelihood examined based on the discriminability of local dynamic features, transitioning from blue (low) to yellow (high). The red boxes outline the precisely determined final anomaly regions, obviously more accurately encapsulate the target anomalies. Green and red dots represent positive and negative click prompts, respectively.}
    \label{fig:compare_result2}
\end{figure}

\subsection{Anomaly Region Identification}

In subsurface anomaly detection, the primary step involves identifying and localizing anomalies within collected GPR data, usually given only some non-target GPR data for initialization. 
We first validated Res-SAM's performance on this anomaly region identification tasks, where 20 non-target GPR data frames were randomly chosen from our dataset to aid in detecting anomalies in the remaining data frames. During the detection phase, click prompts are implemented in each GPR data frame, where click prompts within the anomaly region serve as ``positive'' prompts (left-mouse click), and those outside are defined as ``negative'' prompts (right-mouse click). Fig. \ref{fig:compare_result2} provides representative data frames and corresponding anomaly detection results, where green and red dots represent positive and negative click prompts, respectively. The blue and red boxes indicate anomaly regions detected by SAM and Res-SAM. Notably, SAM tends to segment overly large or incomplete regions containing or intersecting with anomalies, but fails to provide precise and complete areas. In contrast, Res-SAM refines SAM’s results and provides a heatmap of anomaly likelihood, enabling accurate anomaly and boundary localization across different prompt settings.

\begin{figure}[htbp]
    \centering
    \includegraphics[width=\linewidth]{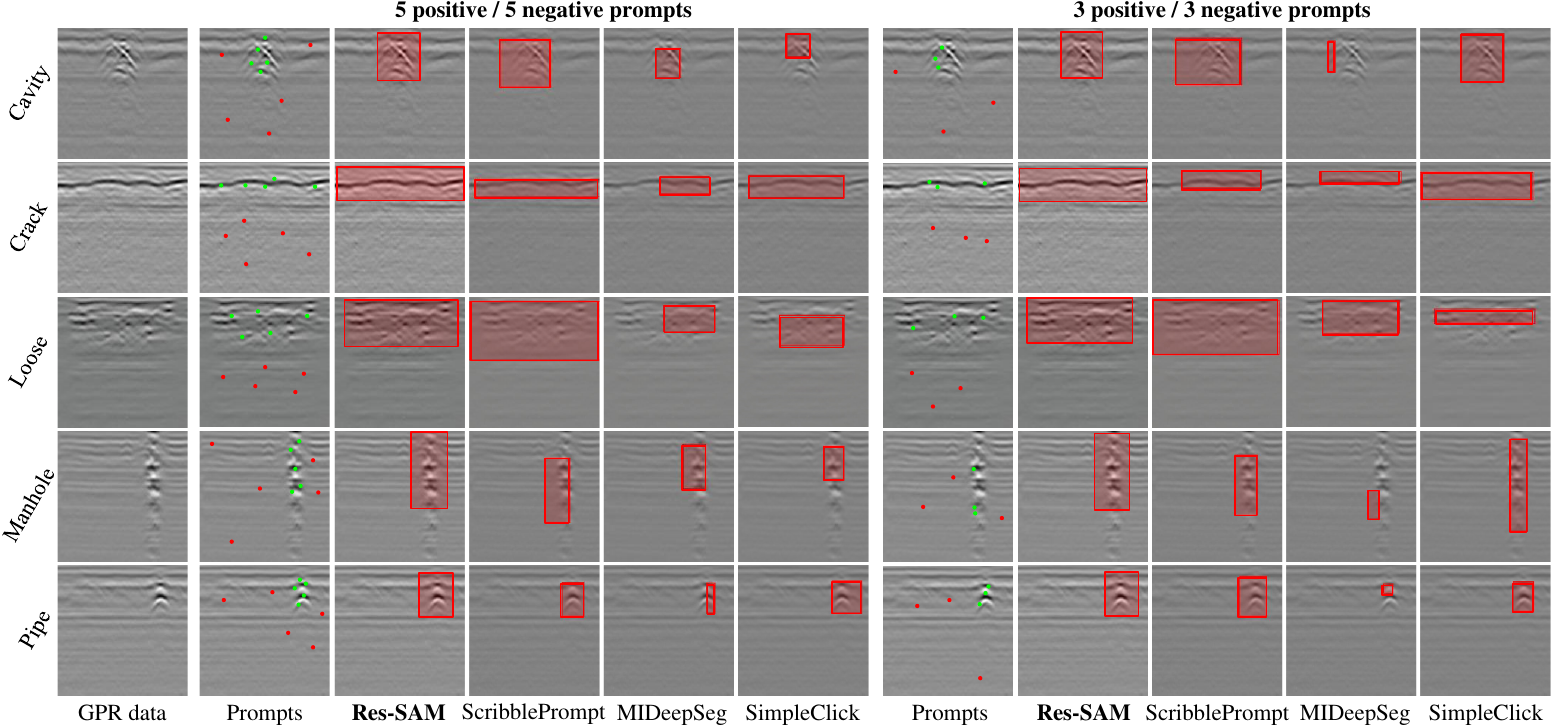}
    \caption{Each row shows an anomaly and the corresponding anomaly regions (the red box) obtained by Res-SAM and three other high-performing methods, supported with two sets of click prompts. }
    \label{fig:compare_result1}
\end{figure}

\begin{figure}[tbp]
    \centering
    % 图例部分
    \begin{minipage}{\linewidth}
        \centering
        \setlength{\fboxsep}{2pt}   % 边框与内容间距
        \setlength{\fboxrule}{0.5pt}  % 边框线宽
        \fbox{\includegraphics[width=0.82\linewidth]{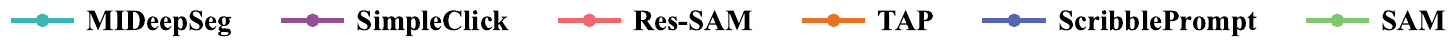}}
    \end{minipage}\\[0.4em] % 添加一些垂直间距
    % 子图部分
    \begin{subfigure}[b]{0.48\textwidth}
        \centering
        \includegraphics[height=1.6in]{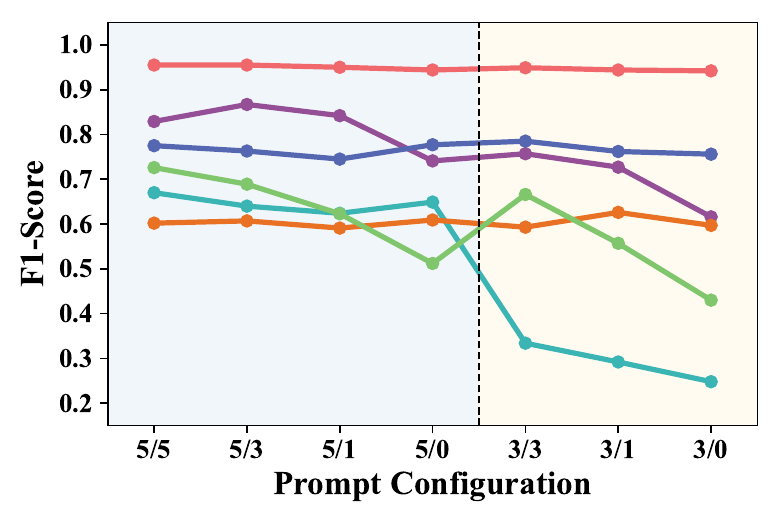}
    \end{subfigure}
    \begin{subfigure}[b]{0.48\textwidth}
        \centering
        \includegraphics[height=1.6in]{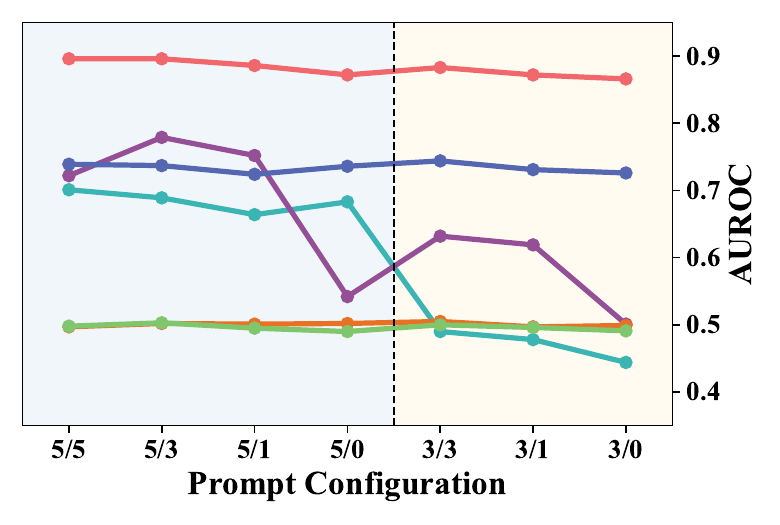}
    \end{subfigure}
    
    \caption{Line chart visualizing the comparative performance of Res-SAM and baseline methods in terms of F1-Score and AUROC across diverse click prompt configurations.}
    \label{fig:f1_auroc_show}
\end{figure}

To the best of our knowledge, there is a lack of research specifically focused on anomaly detection in GPR data. For evaluation, we compared Res-SAM with six state-of-the-art interactive segmentation methods under different prompt settings, including Tokenize Anything via Prompting (TAP) \cite{tap}, SimpleClick \cite{simpleclick}, MIDeepSeg \cite{mideepseg}, ScribblePrompt \cite{scribble}, as well as the original SAM \cite{kirillov2023segment}. These tools were employed under the same prompts as Res-SAM, with the anomalous regions delineated by bounding boxes. Some intuitive comparative examples are given in Fig. \ref{fig:compare_result1}. A detection is considered correct if its IoU with the ground truth exceeds a threshold of 0.5. Table \ref{tab:ad_results} presents AUC and F1-score of the proposed framework and competing methods under different prompt settings.

According to the outcomes and visual evidence, Res-SAM consistently outperforms competing methods across all prompt settings, achieving more precise and comprehensive anomaly identification. Notably, it attains an Area Under the ROC Curve (AUC) of 0.896 and an F1-score of 95.5\% in both the 5/5 and 5/3 settings. Even in the most challenging 3/0 case, Res-SAM maintains an AUC of 0.866 and an F1-score of 94.2\%, highlighting strong generalization. This can be attributed to its ability to exploit both the visual discernibility and wave-collection properties of GPR data. In contrast, purely visual-based methods find anomalies with prompts but fail to extract the regions comprehensively and accurately, leading to significant drops in AUC and F1 scores. Among these methods, SimpleClick performs best in the 5/3 setting with an AUC of 0.779 and an F1-score of 86.7\%. However, as prompts decrease to 3/0, it suffers a sharp decline to 0.501 AUC and 61.6\% F1-score. Similar performance drops occur in other competing methods, indicating the limitations of interactive segmentation under minimal prompt scenarios, whereas Res-SAM consistently excels. Our approach leverages both SAM to approximate anomaly locations and 2D-ESN to refine detection by assessing local changing information in GPR data. Adequately capturing local dual-directional changing information via 2D-ESN fitting, features derived from abnormal patches significantly differ from the established feature bank, resulting in a substantially higher anomaly likelihood (evident as the heatmap in Fig. \ref{fig:compare_result2}) and enabling precise refinement of anomaly regions.

\begin{table*}[tpb]
    \centering
    \caption{Comparison of Res-SAM with baselines in terms of AUROC and F1-Score (F1), tested across various point prompt configurations,
    where ``Pos'' and ``Neg'' represent the number of positive and negative click prompts, respectively.}
    \setlength{\tabcolsep}{2pt}
    \setlength{\aboverulesep}{0pt}
    \setlength{\belowrulesep}{0pt}
    \resizebox{\linewidth}{!}{
    \begin{tabular}{ccccccccccccccccccccc}
    \toprule
    \rowcolor{tabhead}
      & \multicolumn{2}{c}{\textbf{5/5}} & & \multicolumn{2}{c}{\textbf{5/3}} & & \multicolumn{2}{c}{\textbf{5/1}} & & \multicolumn{2}{c}{\textbf{5/0}} & & \multicolumn{2}{c}{\textbf{3/3}} & & \multicolumn{2}{c}{\textbf{3/1}} & & \multicolumn{2}{c}{\textbf{3/0}} \\
      \arrayrulecolor{tabhead} \cmidrule{1-1} \cmidrule{4-4} \cmidrule{7-7} \cmidrule{10-10} \cmidrule{13-13} \cmidrule{16-16} \cmidrule{19-19} \noalign{\vskip -1pt}  \arrayrulecolor{black} \cmidrule{2-3} \cmidrule{5-6} \cmidrule{8-9} \cmidrule{11-12} \cmidrule{14-15} \cmidrule{17-18} \cmidrule{20-21} 
    \rowcolor{tabhead}
    \multirow{-2}{*}{\textbf{Methods}} & \textbf{AUC}   & \textbf{F1}    & & \textbf{AUC}   & \textbf{F1}     & & \textbf{AUC}   & \textbf{F1}  & & \textbf{AUC}   & \textbf{F1}   & & \textbf{AUC}   & \textbf{F1}     & & \textbf{AUC}   & \textbf{F1} & & \textbf{AUC}   & \textbf{F1} \\
    \specialrule{0.8pt}{0pt}{0pt}
    \rowcolor{tabc1}
    MIDeepSeg       & 0.701 & 67.0\% & & 0.689 & 67.0\% & & 0.664 & 62.4\% & & 0.683 & 64.9\% & & 0.490 & 33.4\% & & 0.478 & 29.2\% & & 0.444 & 24.8\%  \\
    % \midrule
    \rowcolor{tabc2}
    SimpleClick     & 0.722 & 82.9\% & & 0.779 & 86.7\% & & 0.752 & 84.2\% & & 0.542 & 74.1\% & & 0.632 & 75.7\% & & 0.619 & 72.7\% & & 0.501 & 61.6\%  \\
    % \midrule
    % \rowcolor{tabc1}
    % FastSAM         & 0.488 & 45.4\% & & 0.352 & 42.8\% & & 0.260 & 34.9\% & & 0.245 & 31.7\% & & 0.498 & 46.4\% & & 0.312 & 38.2\% & & 0.218 & 27.1\%  \\
   % \midrule
    \rowcolor{tabc1}
    TAP             & 0.497 & 60.2\% & & 0.502 & 60.7\% & & 0.501 & 59.1\% & & 0.502 & 60.9\% & & 0.505 & 59.3\% & & 0.497 & 62.6\% & & 0.499 & 59.7\%  \\
  %  \midrule
    \rowcolor{tabc2}
    ScribblePrompt  & 0.739 & 77.5\% & & 0.737 & 76.3\% & & 0.724 & 74.5\% & & 0.736 & 77.7\% & & 0.744 & 78.5\% & & 0.731 & 76.2\% & & 0.726 & 75.6\%  \\
 %   \midrule
    \rowcolor{tabc1}
    SAM             & 0.498 & 72.6\% & & 0.503 & 68.9\% & & 0.495 & 62.3\% & & 0.490 & 51.2\%  & & 0.500 & 66.6\%  & & 0.496 & 55.7\%  & & 0.491 & 43.0\%   \\
    \specialrule{0.8pt}{0pt}{0pt}
    \rowcolor{tabc2}
    \textbf{Res-SAM}         & \textbf{0.896} & \textbf{95.5}\% & & \textbf{0.896} & \textbf{95.5}\% & & \textbf{0.886} & \textbf{95.0}\% & & \textbf{0.872} & \textbf{94.4}\% & & \textbf{0.883} & \textbf{94.9}\% & & \textbf{0.872} & \textbf{94.4}\% & & \textbf{0.866} & \textbf{94.2}\%  \\
    \bottomrule
    \end{tabular}}
  \label{tab:ad_results}
\end{table*}

\subsection{Anomaly Categorization}

\begin{wrapfigure}{r}{0.5\linewidth}
    \centering
     \vspace{-2em}
    {
    \centering
    \begin{minipage}{\linewidth}
        \centering
        \includegraphics[width=0.92\linewidth]{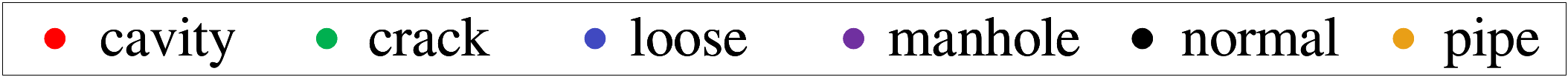}
    \end{minipage}
    }\\[0.4em]
    \centering
	\begin{subfigure}[b]{0.45\linewidth}
        \centering
        \includegraphics[width=\linewidth]{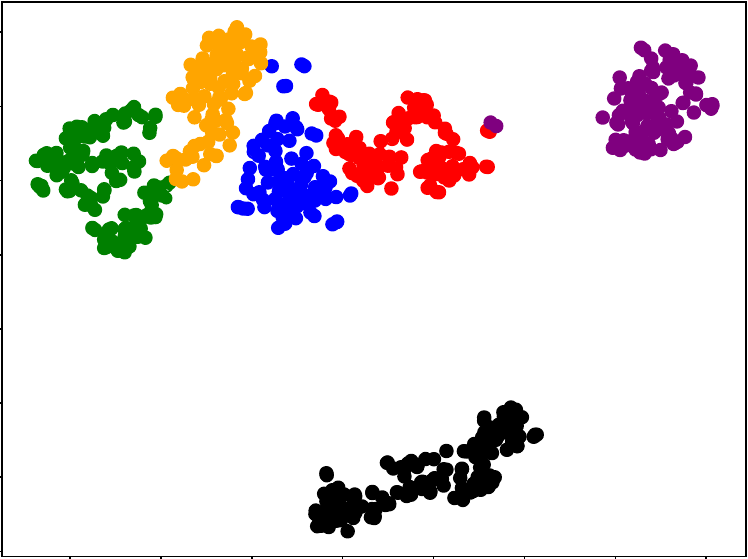}
        \caption{\textbf{Res-SAM}}
    \end{subfigure}
    \begin{subfigure}[b]{0.45\linewidth}
        \centering
        \includegraphics[width=\linewidth]{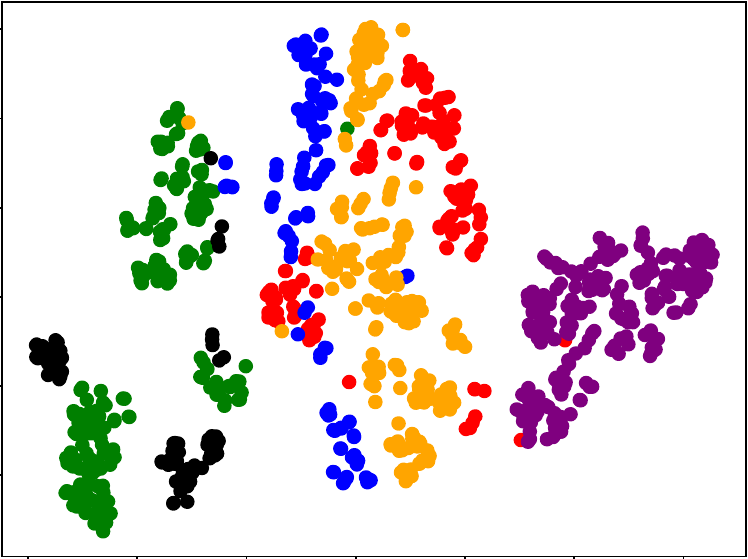}
        \caption{VGG}
    \end{subfigure}
    \begin{subfigure}[b]{0.45\linewidth}
        \centering
        \includegraphics[width=\linewidth]{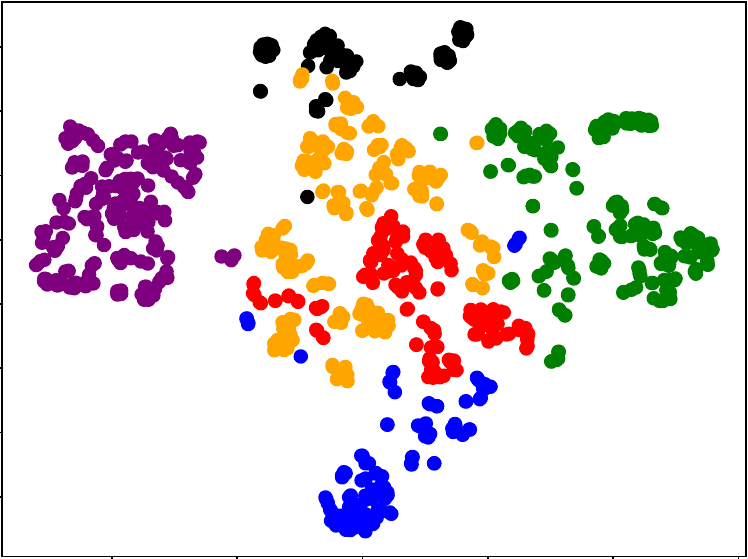}
        \caption{ResNet}
    \end{subfigure}
    \begin{subfigure}[b]{0.45\linewidth}
        \centering
        \includegraphics[width=\linewidth]{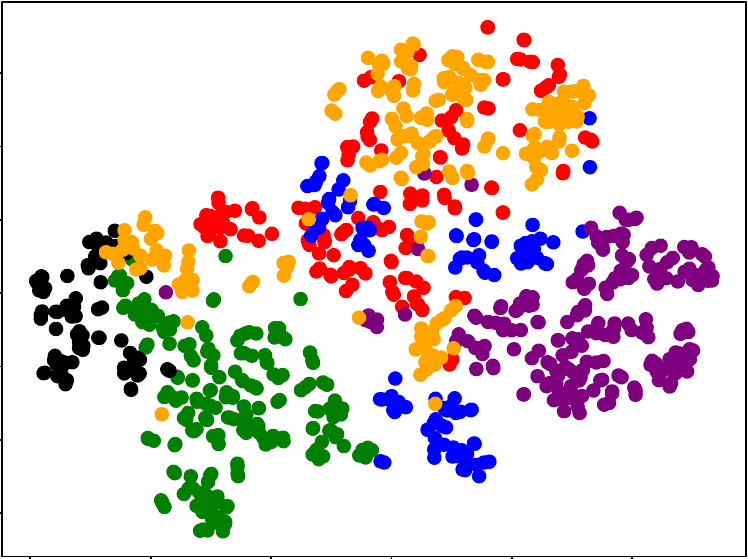}
        \caption{Inception}
    \end{subfigure}
    \begin{subfigure}[b]{0.45\linewidth}
        \centering
        \includegraphics[width=\linewidth]{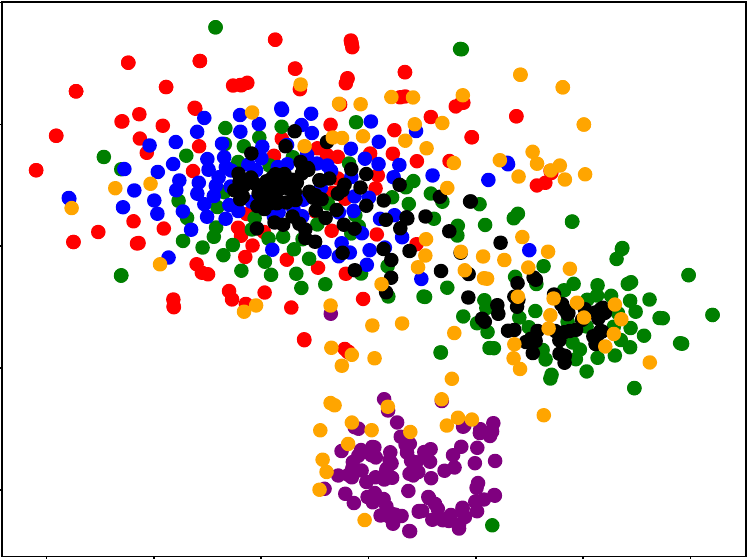}
        \caption{HOG}
    \end{subfigure}
    \begin{subfigure}[b]{0.45\linewidth}
        \centering
        \includegraphics[width=\linewidth]{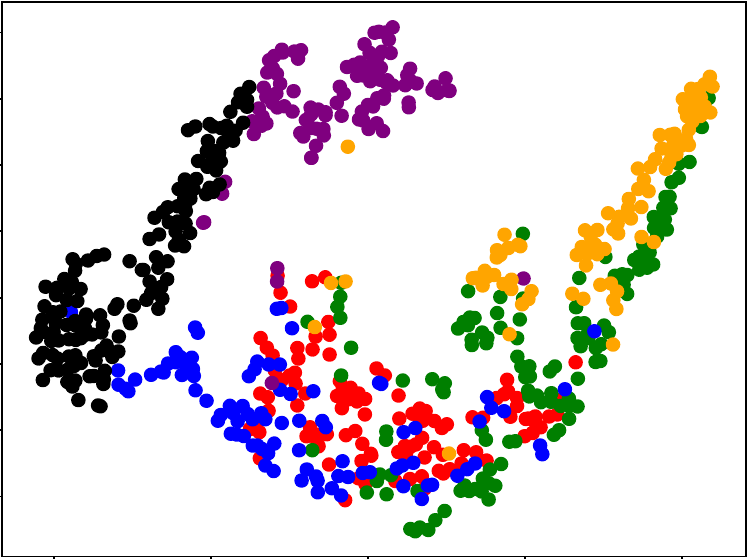}
        \caption{LBP}
    \end{subfigure}
	\caption{Feature distribution via t-SNE. Each point refers to a feature extracted from an anomaly region. Colors indicate anomaly types, with ``normal'' representing features extracted from non-target data.}
	\label{fig:clustering_show}
    \vspace{-2.5em}
\end{wrapfigure}

Upon identifying anomaly regions, categorizing these regions into specific types aids in subsequent repair. Res-SAM allows 2D-ESN to fit these regions, generating respective feature representations, which are then grouped using clusters to categorize anomaly types. This is a one-shot process and avoids offline iterative training. To evaluate anomaly categorization performance, we employ image feature extractors such as Histogram of Oriented Gradients (HOG) \cite{HOG} and Local Binary Pattern (LBP) \cite{LBP}, along with deep learning approaches including VGG \cite{VGG}, ResNet \cite{ResNet}, Inception \cite{Inception}, and Vision Transformer (ViT) \cite{ViT} pre-trained on ImageNet \cite{ImageNet} dataset. Features extracted by these methods are processed using three widely adopted clustering algorithms: K-Means \cite{krishna1999genetic}, Agglomerative Clustering (AC) \cite{gowda1978agglomerative}, and Fuzzy C-Means (FCM) \cite{bezdek1984fcm}. Detailed clustering results are listed in Table \ref{tab:clustring}, while Fig. \ref{fig:clustering_show} visualizes 2D-projected features using t-SNE \cite{van2008visualizing} for a more intuitive comparison.

The results demonstrate that Res-SAM excels in anomaly clustering, outperforming other feature extraction methods across all three clusters. Specifically, Res-SAM combined with Agglomerative Clustering achieves the best metrics, with an accuracy of 0.91, an Adjusted Rand Index (ARI) of 0.85, and a Normalized Mutual Info (NMI) of 0.89, indicating its ability to categorize most underground anomalies correctly. In contrast, current approaches never exceed 0.8 in any clustering metrics. Visually, features derived from Res-SAM exhibit an obvious category separation effect in the projected space, whereas other methods show weaker or even chaotic distributions. Due to the existence of underground anomalies, distinctive dynamics of GPR data arise from variations among and along the collected EM waves. Pre-trained DL networks optimized for visual features struggle to capture these characteristics effectively. Moreover, constructing diverse and accurately labeled training datasets for a specific target area is often impractical. Res-SAM addresses this challenge by focusing on data-inherent dynamics. By establishing connections in both vertical and horizontal directions within GPR data, it comprehensively captures dual-directional dynamics, resulting in category-discriminative features that enhance anomaly clustering accuracy.

\begin{table*}[tbp]
    \centering
    \small
    \caption{Model/feature clustering: Accuracy (Acc), ARI (Adjusted Rand Index), and NMI (Normalized Mutual Info).}
    \setlength{\tabcolsep}{5.pt}
    \setlength{\aboverulesep}{0pt} % 减少表格线上的空白
    \setlength{\belowrulesep}{0pt}
    \resizebox{0.85\linewidth}{!}{
    \begin{tabular}{lccccccccccc}
    \toprule
    \rowcolor{tabhead}
      & \multicolumn{3}{c}{\textbf{K-Means}} & & \multicolumn{3}{c}{\textbf{AC}} & & \multicolumn{3}{c}{\textbf{FCM}} \\
      \arrayrulecolor{tabhead} \cmidrule{1-1} \cmidrule{5-5} \cmidrule{9-9} \noalign{\vskip -1pt}  \arrayrulecolor{black} \cmidrule{2-4} \cmidrule{6-8} \cmidrule{10-12} 
    \rowcolor{tabhead}
    \multirow{-2}{*}{\textbf{Methods}} & \textbf{Acc}   & \textbf{ARI}   & \textbf{NMI}    & & \textbf{Acc}   & \textbf{ARI}   & \textbf{NMI}    & & \textbf{Acc}   & \textbf{NMI}   & \textbf{NMI}  \\
    \specialrule{0.8pt}{0pt}{0pt}
    % HOG & 0.63 & 0.61 & 0.55 & & 0.63 & 0.40 & 0.54 &  & 0.57 & 0.44 & 0.52 \\
    % HOG & 0.50 & 0.26 & 0.55 & & 0.63 & 0.40 & 0.54 &  & 0.57 & 0.44 & 0.52 \\
    \rowcolor{tabc1}
    HOG & 0.43 & 0.21 & 0.44 & & 0.44 & 0.23 & 0.49 &  & 0.35 & 0.17 & 0.28 \\
    \midrule
    \rowcolor{tabc2}
    LBP & 0.54 & 0.37 & 0.48 & & 0.59 & 0.46 & 0.55 &  & 0.46 & 0.32 & 0.45 \\
    \midrule
    \rowcolor{tabc1}
    VGG & 0.71 & 0.67 & 0.76 & & 0.69 & 0.62 & 0.77 &  & 0.76 & 0.74 & 0.75 \\
    \midrule
    \rowcolor{tabc2}
    ResNet & 0.74 & 0.60 & 0.70 & & 0.74 & 0.67 & 0.67 &  & 0.68 & 0.66 & 0.62 \\
    \midrule
    % Inception & 0.47 & 0.29 & 0.40 & & 0.56 & 0.35 & 0.47 &  & 0.41 & 0.29 & 0.31 \\
    \rowcolor{tabc1}
    Inception & 0.58 & 0.36 & 0.49 & & 0.68 & 0.58 & 0.65 &  & 0.30 & 0.09 & 0.13 \\
    \midrule
    \rowcolor{tabc2}
    ViT  & 0.68 & 0.62 & 0.67 & & 0.76 & 0.60 & 0.64 &  & 0.74 & 0.64 & 0.64 \\
    % \midrule
    \specialrule{0.8pt}{0pt}{0pt}
    \rowcolor{tabc1}
    \textbf{Res-SAM} & \textbf{0.83} & \textbf{0.78} & \textbf{0.73} & & \textbf{0.91} & \textbf{0.85} & \textbf{0.89} &  & \textbf{0.90} & \textbf{0.84} & \textbf{0.85} \\
    \bottomrule
    \end{tabular}}
  \label{tab:clustring}
\end{table*}

\section{Discussion}\label{sec3}

With rapid urbanization and infrastructure expansion, subsurface hazards such as cavities, cracks, and structural voids increasingly pose severe risks to urban safety. Effective detection and accurate classification of these hidden anomalies are critical to preventing accidents and ensuring infrastructure stability. GPR has become a widely adopted non-destructive method for this purpose. However, efficient and reliable anomaly detection using GPR still faces several significant challenges. First, labeled anomaly samples are scarce, making datasets highly imbalanced and dominated by non-target samples. Second, the high variability of subsurface environments results in substantial differences in electromagnetic wave characteristics, limiting the generalization ability of existing methods. Third, underground anomalies typically appear as gradual in EM wave reflections rather than distinct boundaries, complicating precise and complete localization via visual-based approaches.

This study introduces the Reservoir-enhanced Segment Anything Model (Res-SAM) framework, designed for efficient and precise detection and classification of subsurface anomalies in GPR data. Compared to conventional methods, Res-SAM combines the visual discernibility of target anomaly with its local wave changing information, providing complete and accurate anomaly regions. Res-SAM incorporates minimal human click prompts, significantly reducing manual labor while enhancing detection reliability and avoiding unnecessary operations on a large volume of easily distinguishable normal data. By capturing and distinguishing dynamic variations within and between EM waves in local GPR data, Res-SAM further refines the identification, substantially improving the completeness and accuracy of output anomaly region. Experimental results clearly demonstrate that Res-SAM consistently outperforms existing state-of-the-art methods. Notably, Res-SAM maintains remarkable stability and accuracy even under minimal interactive prompts, attributed to its effective utilization of intrinsic dynamic changes within local anomaly-associated GPR data. Remarkably, this study highlights the strong generalization capabilities of Res-SAM, which requires only minimal non-target GPR data for initialization, facilitating rapid deployment in unexplored areas and significantly reducing development and application cycles. This adaptability underscores its potential practical value, particularly for urban subsurface infrastructure safety monitoring.

Our research also acknowledges certain limitations worth addressing in future studies. Specifically, while Res-SAM greatly reduces manual intervention, verifying anomaly detection results remains challenging. For anomalies not initially indicated by human prompts, it is essential to implement rapid yet comprehensive background verification processes that complement Res-SAM results. Efficient methods to automate this validation, potentially running in parallel or in the background, would be highly beneficial for practical deployment. Additionally, future work would also explore multi-source data fusion or joint diagnostics using additional sensor data that would further enhance the reliability and accuracy of subsurface anomaly detection.

Overall, the application results of Res-SAM in the field of subsurface anomaly detection are highly encouraging. It significantly reduces labeling costs and improves detection accuracy and efficiency, providing crucial technical support for ensuring the safety of urban infrastructure.

%%局限性
%%减少人为工作量（可以用图）

\section{Methods}\label{sec4}
As depicted in Fig. \ref{fig:framework}, Res-SAM operates in two phases: 
\begin{itemize}
    \item \textbf{Feature Collection:} Using a sliding window, we extract patches on non-target GPR data and individually fit each patch using 2D-ESN to capture their inherent dual-directional changing information, resulting in a set of dynamic features. These features, referred to as ``normal features'' that comprehensively capture local dynamics within non-target GPR data, are stored in a ``Feature Bank''.
    \item \textbf{Anomaly Detection:} Given click prompts, target anomaly's candidate region is firstly extracted. For each point within this region, a centered patch is extracted and fitted with 2D-ESN. The derived dynamic features are compared against the feature bank, measuring a anomaly likelihood for each. Patches with high anomaly likelihood are then merged to accurately outline the refined anomaly region. This process is followed by another application of 2D-ESN fitting to these anomaly regions, enabling clustering of anomalies by type on category-specific dynamic features.
\end{itemize}

In Res-SAM, fitting GPR data patches and capturing its dual-directional dynamics using 2D-ESN form the foundation for data representation, anomaly detection, and further categorization of the extracted features. Thus, this Section begins with an overview of 2D-ESN, followed by explanations of the ``Feature Collection'' and ``Anomaly Detection'' phases.

\subsection{Dual-Directional Echo State Network (2D-ESN)}

Similar to typical reservoir computing netorks\cite{li2012chaotic,chen2014cognitive,yan2024emerging,ehlers2025stochastic}, i.e., the Echo State Network (ESN) \cite{jaeger2001echo}, 2D-ESN comprises an input layer, a hidden layer, and an output layer, as shown in Fig. \ref{2DESN_structure}. Distinctively, 2D-ESN incorporates two reservoirs within the hidden layer to effectively capture dynamics in both horizontal and vertical directions, establishing dual-directional correlations between data points. Each reservoir contains neurons interconnected through fixed and randomly determined connections.

\subsubsection{Data Fitting via 2D-ESN}

\begin{figure}[tbp]
    \centering
    % 第一个子图
    \begin{subfigure}[b]{0.45\textwidth}
        \centering
        \includegraphics[height=0.88in]{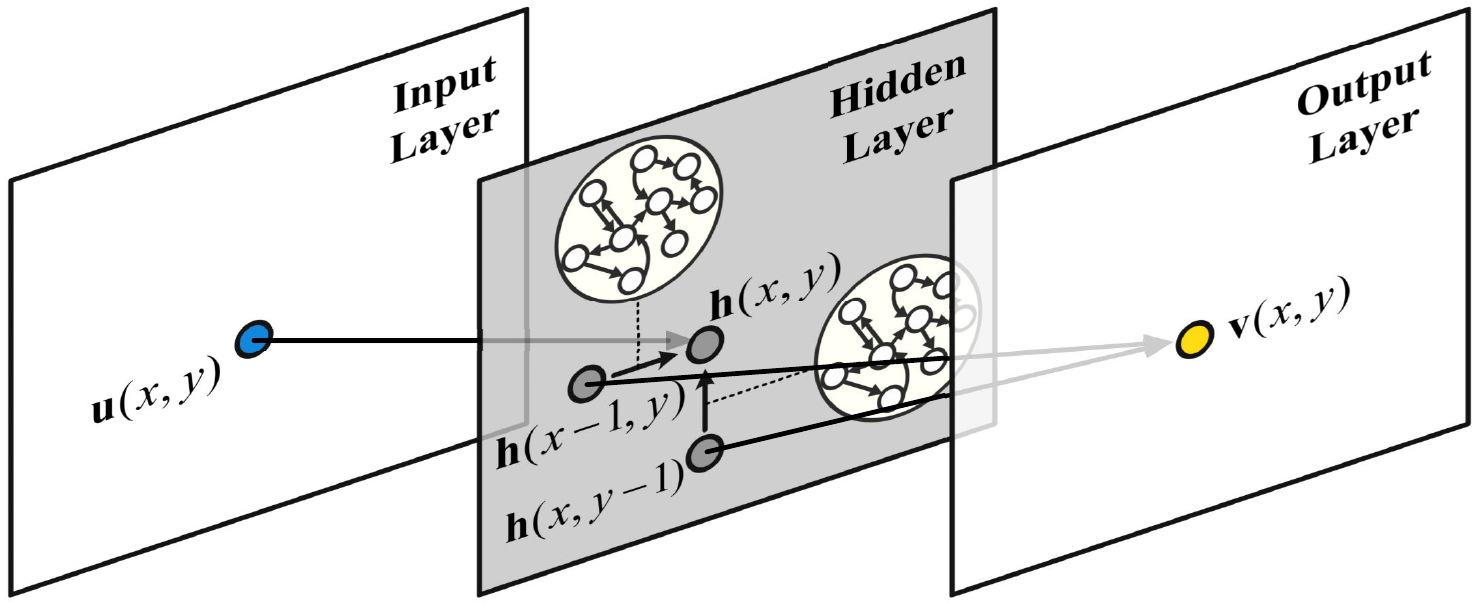}
        \caption{2D-ESN Structure}
        \label{2DESN_structure}
    \end{subfigure}
    \hfill
    % 第二个子图
    \begin{subfigure}[b]{0.25\textwidth}
        \centering
        \includegraphics[height=0.88in]{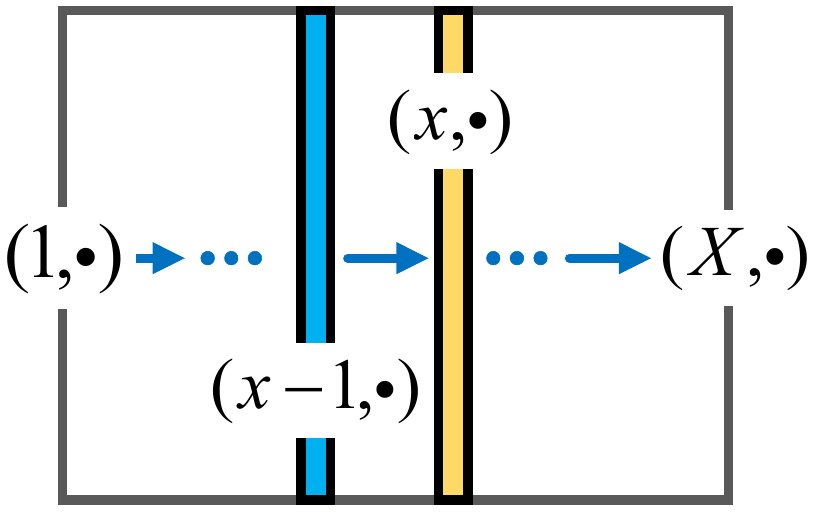}
        \caption{ESN Iteration}
        \label{ESN_iteration}
    \end{subfigure}
    \hfill
    % 第三个子图
    \begin{subfigure}[b]{0.25\textwidth}
        \centering
        \includegraphics[height=0.88in]{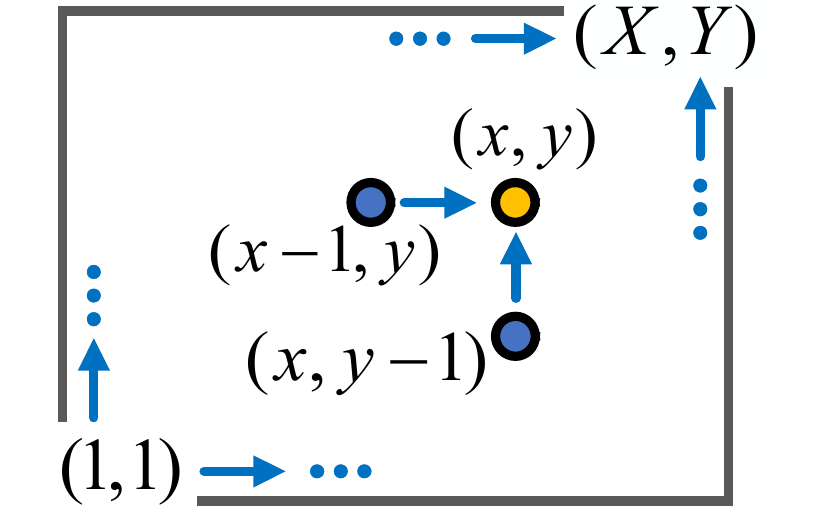}
        \caption{2D-ESN Iteration}
        \label{2DESN_iteration}
    \end{subfigure}
    \caption{\textbf{(a)} 2D-ESN comprises input, hidden, and output layers. The hidden state $\mathbf{h}(x,y)$ is derived from input $\mathbf{u}(x,y)$ and neighboring hidden states $\mathbf{h}(x-1,y)$, $\mathbf{h}(x,y-1)$ from both horizontal and vertical directions. Output $\mathbf{v}(x,y)$ is similarly generated from these adjacent states. \textbf{(b)} ESN iterates unidirectionally, processing each column sequentially with the previous hidden state. \textbf{(c)} 2D-ESN features dual-directional iteration, integrating adjacent processed points in both directions (Eq. \eqref{iteration}).}
\end{figure}

In representing GPR data, we denote it as $\mathbf{u} \in \mathbb{R}^{X\times Y}$, where each point within is localized by coordinates $(x, y)$ and the data value is recorded by $\mathbf{u}(x, y)$. 
As depicted in Fig. \ref{ESN_iteration} and \ref{2DESN_iteration}, 2D-ESN departs from traditional ESN processing by considering correlations in two directions, starting at $(1, 1)$ and progressing through to $(X, Y)$. Data value at each point is sequentially fed into the hidden layer, with its hidden state calculated as: 
\begin{equation}
	\label{iteration}
	\begin{aligned}
		\mathbf{h}(x, y) = g(&\mathbf{W}^{x}\mathbf{h}(x-1,y) + \mathbf{W}^{y}\mathbf{h}(x,y-1) \\
        & + \mathbf{W}^{in}\mathbf{u}(x,y)), 
	\end{aligned}
\end{equation}
where $g$ represents the activation function $tanh$; $\mathbf{W}^{x}$ and $\mathbf{W}^{y}$ are the fixed, randomly determined reservoir weights that describe neuron connections in the reservoir, fulfilling the Echo State Property \cite{buehner2006tighter,yildiz2012re} with a spectral radius between 0 and 1; $\mathbf{W}^{in}$ is the input weight, randomly assigned fixed values from -1 to 1. The initial hidden state $\mathbf{h}(0, y) = \mathbf{h}(x, 0) = 0$ for all $x$ and $y$.

During the iterative process of hidden states defined in Equation \eqref{iteration}, the hidden state of each data point is calculated using the current data point's value and the hidden states of its immediate horizontal and vertical predecessors. This computation integrates both dual-directional predecessors to build correlations between adjacent hidden states, as illustrated in Fig. \ref{2DESN_structure}. As this process unfolds, it establishes and maintains connections between each data point and those previously processed in both directions, ensuring the retention of dual-directional dynamics within the data.

Once the hidden states for all data points are computed, the output layer estimates the output value $\mathbf{v}$ for each point by utilizing the preceding hidden states as follows: 
\begin{equation}
\label{prediction}
\mathbf{v}(x,y) = \mathbf{W}^{out}\mathbf{\hat{h}}(x,y) + \mathbf{a},
\end{equation}
where $\mathbf{W}^{out}$ represents the output weights, $\mathbf{\hat{h}}(x,y) = [\mathbf{h}(x-1, y); \mathbf{h}(x, y-1)]$ aggregates the hidden states from the directly adjacent points, and $\mathbf{a}$ denotes the bias vector.

The 2D-ESN's fitting process is accomplished by ``next point prediction'' \cite{chen2013learning}. This task aims to predict the data value of the next data point using already processed ones, that is, to connect hidden states with corresponding input data values, as illustrated in Fig. \ref{2DESN_structure}. 
It ensures that the output value $\mathbf{v}$, derived from previous hidden states, closely matches the input data value $\mathbf{u}$.
In this task, only the output weight $\mathbf{W}^{out}$ and  bias vector $\mathbf{a}$ need to be solved via ridge regression:
\begin{equation}
	\label{eq2}
        \begin{bmatrix}
        \mathbf{W}^{out}~~\mathbf{a}
        \end{bmatrix}^\text{T}
	= (\mathbf{\tilde{H}}^\text{T}\mathbf{\tilde{H}} + \lambda^2\mathbf{I})^{-1}\mathbf{\tilde{H}}^\text{T}\mathbf{U}, 
\end{equation}
where \(\mathbf{\tilde{H}}\) 
denotes an augmented matrix of hidden states  $\mathbf{\hat{h}}(x,y)$ obtained by collecting hidden states row-wise into a matrix, further expanded by a column of ones to accommodate the bias terms. $\mathbf{U}$ is derived from vectorizing the input values corresponding to each hidden state, \(\mathbf{I}\) represents the identity matrix, and \(\lambda\) serves as a regularization.

Through the above-described fitting process, the dual-reservoir architecture and unique hidden-state iteration of 2D-ESN adequately map the correlations between adjacent points within the data, capturing the data-inherent dual-directional dynamics. Since the output layer plays a pivotal role in mediating these correlations, the derived weights $[ \mathbf{W}^{out}~~\mathbf{a}]$ serve as the dynamic feature of the original GPR data.

The presence of underground anomalies leads to different intra-wave and inter-wave changing information in the collected GPR data, that is, different dual-directional dynamics. Therefore, the dynamic features obtained by fitting the anomaly data using 2D-ESN significantly differ from those derived from normal data. Such feature discrepancy enables effective anomaly detection and clustering.

\subsection{Feature Collection}

Res-SAM only requires non-target GPR data collected from the detecting area, where a sliding window with size \(X \times Y\) and a stride \(s\) is employed to segment each normal GPR data sample into ``Patches'', as illustrated in Fig. \ref{fig:G1234}a.

Each patch undergoes 2D-ESN fitting independently, deriving the dynamic features. The patching and fitting processes are designed to thoroughly capture the local dual-directional dynamics of the non-target GPR data, supporting the subsequent comparison of dynamics extracted from anomaly-associated local GPR data.

The features derived from non-target data patches are collected into a ``Feature Bank'', denoted as:
\begin{equation}
\label{eq:modelbank}
    \mathcal{M}= \phi(\bigcup \limits_{\mathbf{g}_{i}\in \mathcal{G}} \mathcal{P}(\mathbf{g}_{i})), 
\end{equation}
where $\mathcal{P}$ refers to patch extracting process; $\mathcal{G}$ is the collection of non-target GPR data; $\mathbf{g}_i$ denotes a data sample in $\mathcal{G}$; and $\phi$ signifies the 2D-ESN fitting process.

\subsection{Anomaly Detection}
Given the above feature bank, anomaly detection in subsequent GPR data involves the following steps: 
\begin{itemize}
    \item \textbf{Candidate Region Identification:} Using SAM with click prompts to pinpoint anomaly-associated candidate regions on the GPR data.
    \item \textbf{Patch Fitting for Local Dynamic Capture:} For each point within the candidate region, a localized patch centered on this point is extracted and fitted using 2D-ESN, deriving its corresponding dynamic feature.
    \item \textbf{Feature Discrimination and Patch Merging:} For each derived feature in the above step, search the feature bank to identify the nearest normal feature, evaluating the distance to determine the anomaly likelihood. Patches with exceptionally high anomaly likelihoods (indicating anomalies) are recognized, and overlapping ones are merged to outline the anomaly region.
    \item \textbf{Anomaly Categorizing:} Upon obtaining the anomaly regions in the above step, we reapply 2D-ESN to fit these regions respectively and derive dynamic features. Clustering is then implemented to group the features, with each cluster representing a specific type of anomaly, thereby categorizing these anomalies.
\end{itemize}

\subsubsection{Candidate Region Identification}

The presence of underground anomalies leads to changes along and among EM waves, resulting in visible grayscale variations in GPR data as Fig. \ref{abscan}. These visually detectable changes allow for the intuitive and rapid localization of the approximate anomaly region, which SAM, a well-trained segmentation approach, could efficiently handle. However, as previously mentioned, since GPR data essentially collects and represents EM waves, the inherent wave properties such as reflection, diffraction, and attenuation mean that anomalies in GPR data often lack distinct boundaries. Consequently, we recommend using ``click'' prompts, and further providing a rectangular region based on the SAM-segmented area, as illustrated in Fig. \ref{fig:SAM_example}.

\begin{figure}[htbp]
    \centering
    \begin{subfigure}[b]{0.26\textwidth}
        \centering
        \includegraphics[height=1.1in,width=1.1in]{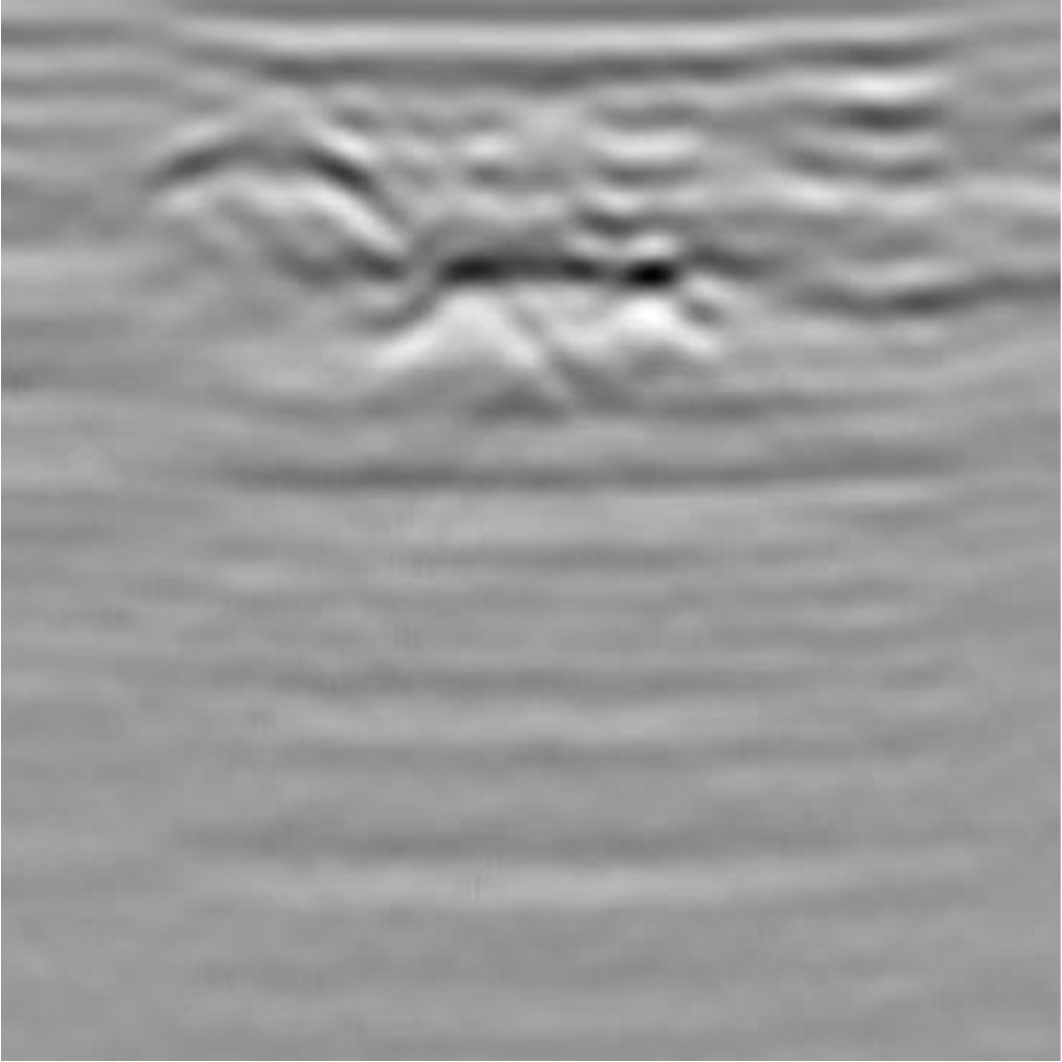}
        \caption{GPR data}
    \end{subfigure}
    \begin{subfigure}[b]{0.26\textwidth}
        \centering
        \includegraphics[height=1.1in,width=1.1in]{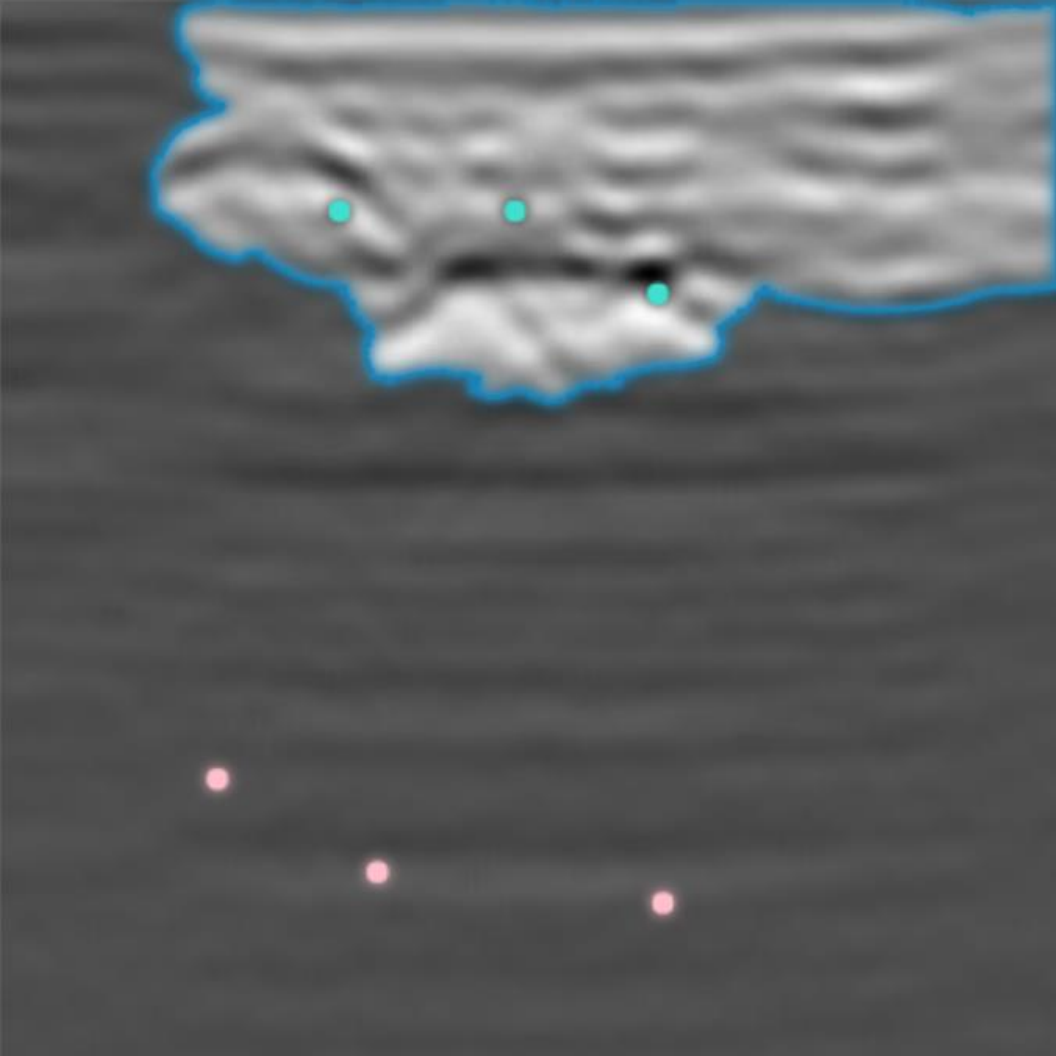}
        \caption{SAM segmentation}
    \end{subfigure}
    \begin{subfigure}[b]{0.26\textwidth}
        \centering
        \includegraphics[height=1.1in,width=1.1in]{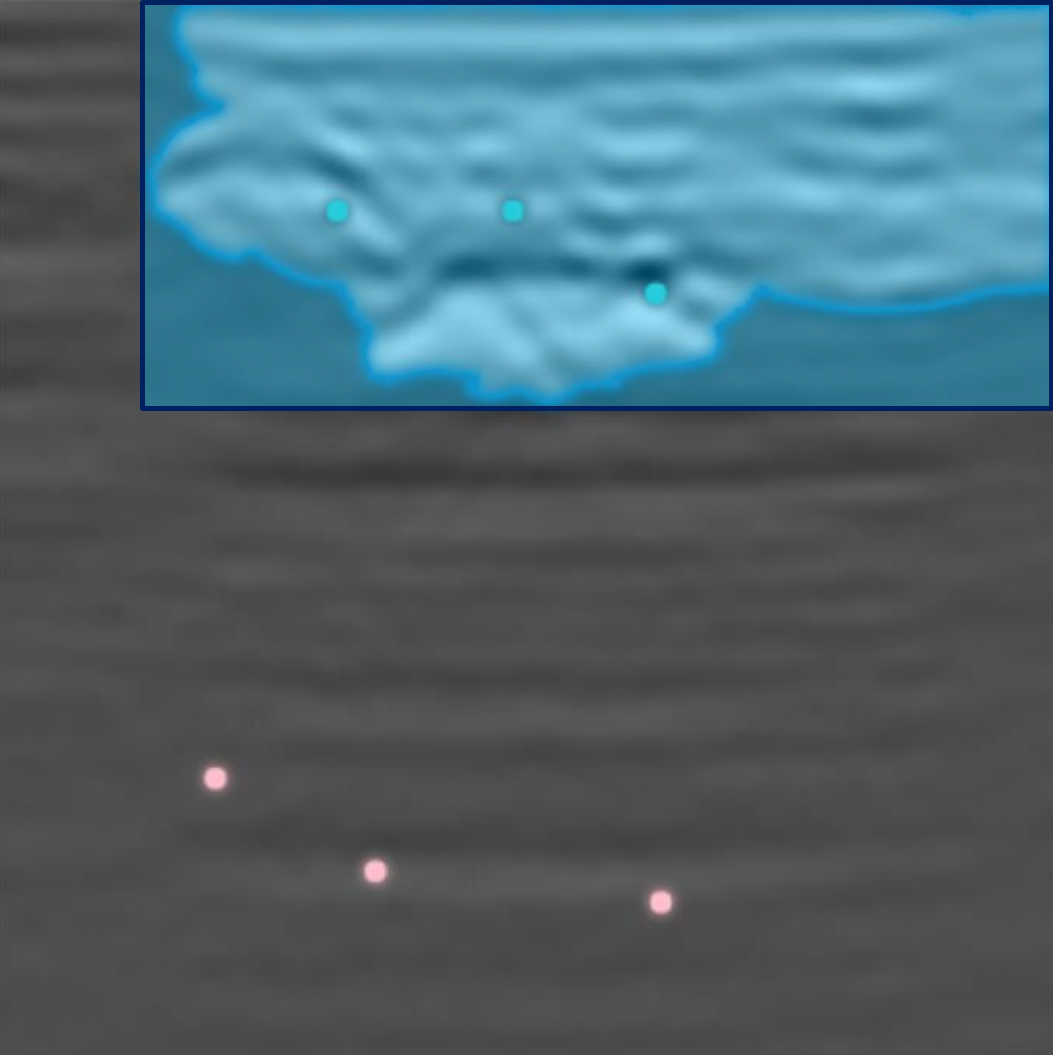}
        \caption{Candidate region}
    \end{subfigure}
    \caption{Supported with simple click prompts, SAM provides the candidate region of the target anomaly. \textbf{(a)} Original GPR data with a visually identifiable anomaly. \textbf{(b)} SAM segmentation of the anomaly based on click prompts. Green/Red points denote click prompts inside/outside the anomaly region. \textbf{(c)} The candidate anomaly region is outlined by the smallest encompassing rectangular box.}
    \label{fig:SAM_example}
\end{figure}

Given an input GPR data sample $\mathbf{g}$ and a set of click prompts $\mathcal{C}=\{(x_i, y_i)\}_{i=1}^{c}$, SAM provides an irregular approximate anomaly-associated region $\widetilde{\mathcal{R}}=\text{SAM}(\mathbf{g}, \mathcal{C})$. However, practical subsurface anomaly detection necessitates displaying the entire anomaly-affected area, typically using rectangular outlines to entirely define these anomaly regions for further analysis. Thus, we use the smallest rectangular box that can encompass the segmented area identified by SAM as our candidate anomaly region, as Fig. \ref{fig:SAM_example} shows, defined by:
\begin{equation}
    \mathcal{R}_{can}=[X_1, X_2] \times [Y_1, Y_2];
\end{equation}
where $X_1$, $X_2$, $Y_1$, and $Y_2$ represent the boundaries of $\widetilde{\mathcal{R}}$:
\begin{equation}
\begin{aligned}
    X_1 = \min_{(x, y) \in \widetilde{\mathcal{R}}} x,& \quad
    Y_1 = \min_{(x, y) \in \widetilde{\mathcal{R}}} y, \\
    X_2 = \max_{(x, y) \in \widetilde{\mathcal{R}}} x,& \quad
    Y_2 = \max_{(x, y) \in \widetilde{\mathcal{R}}} y.
\end{aligned}
\end{equation}
Subsequently, we will analyze the local dynamic changes from this defined candidate region to delineate the target anomaly region more precisely.

\subsubsection{Patch Fitting for Local Dynamic Feature}

While changes in GPR data due to underground anomalies are visually identifiable and could be generally depicted (as shown in Fig. \ref{fig:SAM_example}), GPR data fundamentally consists of wave collections. The variations in local dual-directional dynamics caused by these underground anomalies help describe the abnormal area more accurately.

\begin{figure}[htbp]
	\centering
    \includegraphics[width=\linewidth]{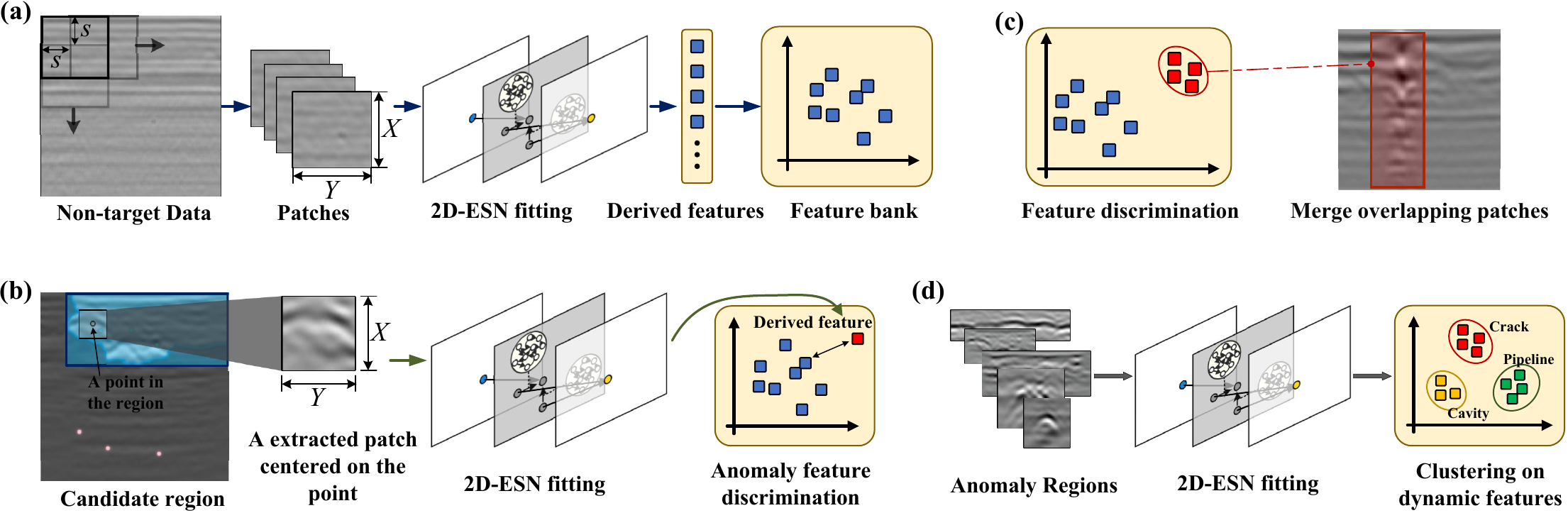}
    \caption{Detailed process of Res-SAM. \textbf{(a)} Non-target GPR data is divided into ``Patches'' using a sliding window. Each patch is fitted by 2D-ESN with the obtained dynamic features collected into a ``Feature Bank.'' \textbf{(b)} Within candidate regions, localized patches around each point are fitted via 2D-ESN to derive dynamic features, of which the ``anomaly likelihood'' is further evaluated based on feature discrimilites. \textbf{(c)} Overlapping patches identified as anomalous are merged into final anomaly regions. \textbf{(d)} Anomaly regions are refitted by 2D-ESN to extract dynamic features. Features of the same anomaly type are clustered together, while dissimilar ones are separated, enabling categorization of distinct anomalies, with each cluster representing a unique type.}
    \label{fig:G1234}
\end{figure}

To capture local dynamics, each point in the candidate region $\mathcal{R}_{can}$ serves as the center for extracting a local patch of size $X \times Y$, as Fig. \ref{fig:G1234}b shows, excluding patches that extend beyond boundaries. Each of these patches is then fitted via 2D-ESN to derive dynamic features, and all such features are accumulated into a set $\mathcal{F}$. 

Underground anomalies lead to distinct local changing information along and among EM waves, deriving dynamic features that set apart from those from non-target data. Consequently, features developed from these anomalous patches exhibit marked differences from those generated from normal data. Specifically, features derived from non-target data tend to cluster closely together, whereas those from anomalies are clearly distinct and positioned significantly farther from the normal features. 

\subsubsection{Feature Discrimination and Patch Merging}

Based on the discrepancies between derived features, we can ascertain whether each patch extracted in the previous step is anomaly-relevant. Patches that are deemed anomaly-relevant are retained; others are discarded. Overlapping patches identified to be anomalies are merged to comprehensively define the final anomaly region.

For each feature $f^{*}\in \mathcal{F}$ in the above step, an anomaly likelihood $\mathcal{S}(f^{*})$ is introduced. Specifically, we search the feature bank $\mathcal{M}$ to identify the nearest normal feature. The anomaly likelihood for $f^{*}$ is defined by:
\begin{equation}
    \label{fun：anomalyscore}
    \mathcal{S}(f^{*}) = \min\limits_{f \in \mathcal{M}} L_{2}(f^{*},f),
\end{equation}
where \(f\) denotes one of the normal features within feature band \(\mathcal{M}\), and \(L_2(\cdot, \cdot)\) represents the $L_2$-norm metric to measure distance between features.

A binary classifier \(\mathcal{H}\) is then employed to discriminate normal (denoted as $1$) and abnormal (denoted as $0$) features:
\begin{equation}
    \mathcal{H}(f^{*}) = \begin{cases}
    0 & \text{if } \mathcal{S}(f^{*}) > \beta, \\
    1 & \text{if } \mathcal{S}(f^{*}) < \beta,
    \end{cases}
\end{equation}
where \(f^{*}\) represents the feature under evaluation, \(\mathcal{S}\) is given in Equation \eqref{fun：anomalyscore}, and \(\beta\) is a predefined threshold.

Throughout the above, patches indicative of anomalies are identified. To accurately define the region, overlapping anomaly-relevant patches are consolidated into the smallest possible rectangular box that encompasses them, as Fig. \ref{fig:G1234}c. This rectangular box serves as the final anomaly region.

\subsubsection{Anomaly Categorizing on Dynamic Features}

Different underground anomalies exhibit distinct changing information within and between the collected EM waves, resulting in unique dynamics. Consequently, dynamic features derived from various types of anomalies show noticeable differences, while features from similar anomalies exhibit similarities due to consistent inherent dynamics. After pinpointing the anomaly regions in the above step, we reapply 2D-ESN to fit these specific regions, deriving corresponding dynamic features. Subsequently, we implement clustering techniques, potentially utilizing methods such as K-Means, Agglomerative Clustering, or Fuzzy C-Means, to group these features, as illustrated in Fig. \ref{fig:G1234}d. Each cluster represents a distinct type of anomaly, thus enabling the categorization of the anomalies identified.

\section{Data Availability}
The authors declare that all relevant data are available in the paper or from the corresponding author on request.

\section{Code Availability}
The code script for the Res-SAM framework, along with a video demonstration guide, is available at \url{https://github.com/zhouxr6066/Res-SAM}.

% \begin{appendices}

% \section{Section title of first appendix}\label{secA1}

% An appendix contains supplementary information that is not an essential part of the text itself but which may be helpful in providing a more comprehensive understanding of the research problem or it is information that is too cumbersome to be included in the body of the paper.

%%=============================================%%
%% For submissions to Nature Portfolio Journals %%
%% please use the heading ``Extended Data''.   %%
%%=============================================%%

%%=============================================================%%
%% Sample for another appendix section			       %%
%%=============================================================%%

%% \section{Example of another appendix section}\label{secA2}%
%% Appendices may be used for helpful, supporting or essential material that would otherwise 
%% clutter, break up or be distracting to the text. Appendices can consist of sections, figures, 
%% tables and equations etc.

% \end{appendices}

%%===========================================================================================%%
%% If you are submitting to one of the Nature Portfolio journals, using the eJP submission   %%
%% system, please include the references within the manuscript file itself. You may do this  %%
%% by copying the reference list from your .bbl file, paste it into the main manuscript .tex %%
%% file, and delete the associated \verb+\bibliography+ commands.                            %%
%%===========================================================================================%%

\bibliography{main}% common bib file
%% if required, the content of .bbl file can be included here once bbl is generated
%%\input sn-article.bbl

\end{document}